# Surface plasmon oscillations in a semi-bounded semiconductor plasma

M. SHAHMANSOURI[1,*], A. P. MISRA[2]

[1]*Department of Physics, Faculty of Science, Arak University, Arak 38156- 8 8349, Iran.*

[2]*Department of Mathematics, Siksha Bhavana, Visva-Bharati University, Santiniketan-731 235, India.*

[*]Corresponding author: mshmansouri@gmail.com

**Abstract**

We study the dispersion properties of surface plasmon oscillations in a semi-bounded semiconductor plasma with the effects of the Coulomb exchange force associated with the spin polarization of electrons and holes as well as the effects of the Fermi degenerate pressure and the quantum Bohm potential. Starting from a quantum hydrodynamic (QHD) model coupled to the Poisson equation, we derive the general dispersion relation for surface plasma waves. Previous results in this context are recovered. The dispersion properties of the surface waves are analyzed in some particular cases of interest and the relative influence of the quantum forces on these waves are also studied for a nano-sized GaAs semiconductor plasma. It is found that the Coulomb exchange effects significantly modify the behaviors of the surface plasmon waves. The present results are applicable to understand the propagation characteristics of surface waves in solid density plasmas.

(Some figures may appear in color only in the online journal)

## 1 Introduction

The propagation of surface waves in plasmas and in conducting solids has been a topic of important research over the last many years [1-12]. Surface Plasmon (SP) waves can be excited due to the collective oscillation of free electrons at the interface of plasmas of different densities or plasma-vacuum interface. The characteristics of these surface modes have been a subject of many experimental and theoretical investigations in plasmas because of their special frequency spectrum [1-3].

Due to the great degree of miniaturization of semiconductors in electronic devices (in which the electron density is high and the temperature is low) the thermal de Broglie wavelength of charge particles can be comparable to the spatial variation of the doping profiles. Thus, the typical quantum effects such as the exchange-correlation, the quantum fluctuation due to the density correlation and the degenerate pressure play a non-negligible role in the electronic components to be constructed in the future.

In the previous investigations [4-22], the basic features of the surface waves in semi-bounded plasmas have been investigated under the influence of e.g., the quantum tunneling [4-20], the relativistic effects [12], the spin of fermions [14,15, 22], the collisional effects [9,11], the nonlocality effects [16], the exchange effects [17-21] as well as the external magnetic field [7]. However, the effect of the external magnetic field on surface plasma waves reported in Ref. 7 and elsewhere has not been properly considered, and some comments in these contexts in can be found in, e.g., Refs. 28-30. On the other hand, when the quantum effects are significant the Coulomb exchange (CE) force plays an important role [6,20,22]. The dispersion properties of electromagnetic surface waves on the quantum plasma half space with the effects of exchange-correlation have been studied [17]. The theory has been advanced in electron-positron plasmas as well [8]. However, the effects of the CE force have not yet been considered before in the context of surface plasma waves.

In the present work, we study the propagation characteristics of surface electromagnetic (SEM) waves under the influence of the Coulomb exchange effects [23] in a semiconductor electron-hole plasma. We employ the QHD model for fluid electrons and holes accounting for the effects of the quantum force associated with the Bohm potential, the degenerate pressure and the CE force. The plasma is immersed in an external uniform magnetic field. However, we consider the external magnetic field to be along the $y$–axis and the wave propagation parallel to it [24], i.e., $\vec{B}_0 // \hat{y} // \vec{k}$, where $k$ is the wavenumber (Fig. 1). With this assumption, the Lorentz force becomes zero and does not have any influence on the wave propagation. Nevertheless, the contribution of the external magnetic field appears through the spin polarization of electrons and holes, which modifies the CE interaction term and the equation of state [20,24]. We then drive the general dispersion relation of SEM waves by the effects of the CE interaction and other quantum forces. It must b added here that, as considered in the previous investigations [25-30], the oblique wave propagation relative to the external magnetic field can be a problem of interest, but is beyond the scope of the present investigation, and is left for future studies.

## 2 Theoretical model

We consider a semiconductor quantum plasma consisting of electrons and holes which fills the half-space $x>0$. The plane $x=0$ forms an interface between the plasma region and vacuum $x<0$ (Fig. 1). In a semiconductor plasma with high conductivity, we suppose that the electron and hole densities are not equal [31]. The QHD equations which model the dynamical behaviors of electrons and holes are given by [8,20]

$$\frac{\partial n_j}{\partial t} + n_{j0}\nabla \cdot \mathbf{v}_j = 0, \tag{1}$$

$$m_j \frac{\partial \mathbf{v}_j}{\partial t} = q_j \mathbf{E} - \frac{1}{n_j}\nabla P_j - \nabla V_{qj} - \nabla V_{cj} + q_j \mathbf{v}_j \times \vec{B}_0 \tag{2}$$

where $n_j$, $n_{j0}$, $\mathbf{v}_j$, $m_j$ and $q_j$ refer to the number density, equilibrium number density, velocity, mass, and charge of $j$-species particle respectively. Also, $\mathbf{E}$ is the electric field to be described later by the Maxwell-Poisson equations; $P_j$, $V_{qj}$ and $V_{cj}$ to be defined in the following. The second term on the right-hand side (RHS) of Eq. (2) refers to the pressure gradient term of which the weakly relativistic degenerate species obey the statistical pressure law [32] $P_j = \vartheta_{3D}(3\pi^2)^{2/3}\hbar^2 n_j^{5/3}/5m_j = m_j v_{Fj}^2 n_j^{5/3}/5n_{j0}^{2/3}$. Here, $v_{Fj} = \sqrt{\vartheta_{3D}}\hbar(3\pi^2 n_{j0})^{1/3}/m_j$ is the Fermi velocity of j-species particle with $j=e(h)$ denoting the electrons (holes) and $\vartheta_{3D}$ is the degree of spin polarization given by $\vartheta_{3D} = [(1+\eta)^{5/3} + (1-\eta)^{5/3}]/2$ with $\eta$ being the polarization defined by $\eta = |n_\uparrow - n_\downarrow|/|n_\uparrow + n_\downarrow|$ (The suffices $\uparrow$ and $\downarrow$ denote the species with spin up and spin down). In the cases of full $(\eta=1)$ and zero $(\eta=0)$ spin polarization, we have $\vartheta_{3D} = 2^{2/3}$ and $\vartheta_{3D} = 1$. The third term on the RHS of Eq. (2) represents the particle dispersion, associated with the density correlation due to the quantum tunneling effect, given by [31,33] $V_{qj} = -\hbar^2 \nabla^2 \sqrt{n_j}/2m_j\sqrt{n_j}$. The fourth term on the RHS of Eq. (2) represents the CE effect [20,23,24,32], given by $V_{cj} = -3(3/\pi)^{1/3}\xi_{3D}^{(j)} e^2 n_j^{4/3}/4$, where $\xi_{3D}$ measures the degree of spin polarization given by $\xi_{3D} = (1+\eta)^{4/3} - (1-\eta)^{4/3}$. In the limiting cases of fully $(\eta=1)$ and zero $(\eta=0)$ spin polarization, $\xi_{3D}$ takes the values of $2^{4/3}$ and $0$, respectively. It turns out that in the case where the particles are not spin polarized, i.e., $(n_\uparrow \approx n_\downarrow)$, then the CE interaction becomes zero and does not contribute to the electron/hole interactions. Finally, the last term in Eq. (2) refers to the Lorentz force.

On the other hand, the electromagnetic fields are governed by the Maxwell-Poisson equations as

$$\nabla \times \mathbf{E} = -\frac{1}{c}\frac{\partial \mathbf{B}}{\partial t}, \qquad (3)$$

$$\nabla \times \mathbf{B} = \frac{1}{c}\frac{\partial \mathbf{E}}{\partial t} - \frac{4\pi e}{c}(n_{e0}\mathbf{v}_e - n_{h0}\mathbf{v}_h), \qquad (4)$$

$$\nabla \cdot \mathbf{B} = 0, \qquad (5)$$

$$\nabla \cdot \mathbf{E} = 4\pi e(n_h - n_e). \qquad (6)$$

In what follows, we split up the physical quantities into their equilibrium and perturbation parts as: $n_j \sim n_{j0} + n_{j1}$, $\mathbf{v}_j \sim 0 + \mathbf{v}_{j1}$, $\mathbf{E} \sim 0 + \mathbf{E}$ and $\mathbf{B} \sim 0 + \mathbf{B}$. Then, using the space-time Fourier transforms

$$f(x,y,t) = \frac{1}{(2\pi)^3}\iint d^3k d\omega F(k,x,\omega)\exp[i(ky - \omega t)] \qquad (7)$$

where $\mathbf{k} = k\{\hat{y}\}$ and the perturbed quantities $f(x,y,t)$ transform into $F(k,y,\omega)$ by Eq. (7), we obtain the following expression for the perturbed density as

$$\omega^2 N_j(x) = s_j \omega_{pj}^2 [N_e(x) - N_h(x)] - (\Lambda_j + \beta_j^2 k^2)(-k^2 + \frac{\partial^2}{\partial x^2})N_j(x) \qquad (8)$$

where $\omega_{pj} = \sqrt{4\pi n_{j0}e^2/m_j}$ is the plasma oscillation frequency, $s_e = +1$, $s_h = -1$, $\Lambda_j = \alpha_j^2 - \gamma_j$, $\alpha_j^2 = v_{Fj}^2$, $\beta_j^2 = \hbar^2/4m_j^2$, and $\gamma_j = \xi_{3D}^{(j)}(3/\pi)^{1/3}e^2 n_{j0}^{1/3}/m_j$. The parameter $\gamma_j$ appears due to the influence of the CE interaction potential. Then after some straightforward algebra and by introducing the wave number $q_E = \{(a_e a_h - \omega_{pe}^2 \omega_{ph}^2)/[a_h \Lambda_e + a_e \Lambda_h]\}^{1/2}$, Eq. (8) can be written in the form of the wave equation as

$$\frac{d^2 N_j(x)}{dx^2} - q_E^2 N_j(x) = 0, \qquad (9)$$

where $a_j = \omega_{pj}^2 + \Delta_j k^2 - \omega^2$, $\Delta_j = v_{Fj0}^2\left[1 - \frac{1}{4}H_j^2(3\xi_{3D}^{(j)}\vartheta_{3D}^{(j)} - \lambda_{Fj}^2 k^2)\right]$, $\lambda_{Fj} = v_{Fj}/\omega_{pj}$, $H_j = \hbar\omega_{pj}/m_j v_{Fj}^2$,

$v_{Fj0} = \hbar(3\pi^2 n_{j0})^{1/3}/m_j$. Here, the dimensionless quantum parameter $H_j$ refers to the ratio of the j-species Plasmon energy to the Fermi energy. In order to obtain Eqs. (8) and (9), we have neglected the very slow nonlocal variations ($k_y^{-2}(\partial^4/\partial x^4) << \partial^2/\partial x^2 << k_y^2$) [6,8,17].

On the other hand, a similar wave equation can be obtained for the magnetic field from Eqs. (2)-(4), as follows

$$\frac{d^2 B(x)}{dx^2} - q_M^2 B(x) = 0, \qquad (10)$$

where $q_M = \{k^2 + (\omega_{pe}^2 + \omega_{ph}^2 - \omega^2)/c^2\}^{1/2}$. Though, there are some possible modes, namely the degenerate or singular modes [34], we are, however, interested in the solutions of Eqs. (9) and (10) of the following forms

$$N_j(x) = \begin{cases} 0, & x < 0 \\ A_j \exp(-q_E x), & x \geq 0 \end{cases} \qquad (11a)$$

$$\mathbf{B}(x) = \begin{cases} \mathbf{C}_1 \exp(q_v x), & x < 0 \\ \mathbf{C}_2 \exp(-q_M x), & x \geq 0 \end{cases} \qquad (11b)$$

where $q_V$ is the limiting value of $q_M$ in vacuum, given by, $q_V = (k^2 - \omega^2 c^2)^{1/2}$, and $A_e$, $A_h$ and $\mathbf{C}_{1,2}$ are constant coefficients to be determined later by using the proper boundary conditions. Next, the Maxwell equation (4) is used to obtain the required solution for the electric field

$$\mathbf{E}(x) = \begin{cases} \dfrac{c}{\omega}(-k\hat{i} - iq_v\hat{j})C_1\exp(q_v x) & x < 0 \\ G\left\{\dfrac{cC_2}{\omega}e^{-q_M x}(-k\hat{i} + iq_M\hat{j}) - \dfrac{4\pi e}{\omega^2}\sum_{e,i} K_j A_j e^{-q_E x}\left[-q_E\hat{i} + ik\hat{j}\right]\right\} & x \geq 0 \end{cases} \qquad (12)$$

where $G = \omega^2/[\omega^2 - \omega_{pe}^2 - \omega_{ph}^2]$ and $K_j = [\Delta_j - \frac{1}{4}v_{Fj0}^2 H_j^2 \lambda_{Fj}^2 q_E^2)]$.

We determine the unknown coefficients $A_e$, $A_h$, $C_1$ and $C_2$ by using the appropriate boundary conditions, i.e., the continuity of the tangential component of the wave electric field, i.e., ($E_y|_{x=0^+} = E_y|_{x=0^-}$) and that of the tangential component of magnetic field, i.e., ($B_z|_{x=0^+} = B_z|_{x=0^-}$) as well as the vanishing of the x-component of electron and hole velocities at $x=0$. Note that the surface modes restrict to that part of solutions which decay away from the interface in both plasma and vacuum regions. Applying the boundary conditions as mentioned and after some straightforward algebra, we obtain the following dispersion relation

$$(\chi_e \omega_{pe}^2 - K_e q_E)(\chi_h \omega_{ph}^2 + K_h q_E) - \chi_e \chi_h \omega_{pe}^2 \omega_{ph}^2 = 0 \tag{13}$$

where $\chi_j = v_{Fj0}^2 G(G(k^2 - q_E q_M) - q_E q_v)[4 - H_j^2(3\xi_{3D}^{(j)} \vartheta_{3D}^{(j)} - \lambda_{Fj}^2(k^2 - q_E^2))]/4\omega^2(q_v + G q_M)$. The dispersion relation (13) is our main result of the present investigation, which is significantly modified by the effects of the CE interaction of electrons and holes.

## 3 Dispersion relation and numerical analysis

To recover some previous known results in the literature, we recast Eq. (13) in a more explicit form as

$$q_E \left(k^2(\omega_{pe}^2 + \omega_{ph}^2) - q_E[q_v(\omega^2 - \omega_{pe}^2 - \omega_{ph}^2) + q_M \omega^2]\right)\left(\omega^2 - \omega_{pe}^2 - \omega_{ph}^2\right) = 0 \tag{14}$$

In absence of the hole species, the first factor in Eq. (14) gives

$$k^2 \omega_{pe}^2 = q_E[q_v(\omega^2 - \omega_{pe}^2) + q_M \omega^2] \tag{15}$$

Disregarding the CE effects from Eq. (15), one can obtain the same dispersion relation as obtained by Lazar *et al* [6] and the similar form in some previous work [17]. On the other hand, in the electrostatic limit, i.e., $c \to \infty$, the first factor in Eq. (14) gives

$$q_E(2\omega^2 - \omega_{pe}^2 - \omega_{ph}^2) = k(\omega_{pe}^2 + \omega_{ph}^2). \tag{16}$$

Furthermore, in the absence of hole species and when the overcritical plasma density limit is considered (i.e., $k^2 \Delta_e << |\omega_{pe}^2 - \omega^2|$), Eq. (16) reduces to

$$\omega = \frac{\omega_{pe}}{\sqrt{2}}\left[1 + \frac{k\sqrt{\Delta}}{\sqrt{2}\omega_{pe}}\right] \tag{17}$$

This dispersion relation agrees with the results of some previous works, viz., Shahmanouri [17], Lazar *et al* [6] in absence of the CE effect, and that of Ritchie [4] in absence of the quantum and CE effects. Also, ignoring the quantum effects, Eq. (17), in limit of cold plasma, yields the frequency of surface plasmons $\omega = \omega_{pe}/\sqrt{2}$.

Furthermore, in the overcritical plasma density limit (i.e., $k^2 \Delta_j << |\omega_{pj}^2 - \omega^2|$) Eq. (16) reduces to the following form

$$\omega = \frac{\sqrt{\omega_{pe}^2 + \omega_{ph}^2}}{\sqrt{2}} \left[ 1 + \frac{1}{\sqrt{2}} \frac{k\zeta \sqrt{\omega_{pe}^2 - \omega_{ph}^2}}{\omega_{pe}^2 + \omega_{ph}^2} \right] \qquad (18)$$

where $\zeta = \sqrt{2v_{Fe0}^2 [1 - 0.25 H_e^2 (3\xi_{3D}^{(e)} \vartheta_{3D}^{(e)} - \lambda_{Fe}^2 k^2)] + 2v_{Fh0}^2 [1 - 0.25 H_h^2 (3\xi_{3D}^{(h)} \vartheta_{3D}^{(h)} - \lambda_{Fh}^2 k^2)]}$. Introducing the dimensionless quantities: $W = \omega/\omega_{pe}$ and $K = k_y v_{Fe0}/\omega_{pe}$, Eq. (18) can be rewritten as

$$W = \frac{\sqrt{1+\delta m}}{\sqrt{2}} \left[ 1 + \frac{1}{\sqrt{2}} \frac{KZ\sqrt{1-\delta m}}{1+\delta m} \right] \qquad (19)$$

where $m = m_e/m_h$, $\delta = n_{h0}/n_{e0}$ and

$Z = \sqrt{2[1 - 0.25 H_e^2 (3\xi_{3D}^{(e)} \vartheta_{3D}^{(e)} - K^2)] + 2\delta^{2/3} [1 - 0.25 \delta^{1/3} m^2 H_e^2 (3\xi_{3D}^{(h)} \vartheta_{3D}^{(h)} - K^2/\delta^{1/3} m)]}$. In order to analyze the dispersion equation (19) numerically, we consider some typical values of the physical parameters for which the electrons and holes are degenerate, viz. $n_{j0} \sim 10^{26} m^{-3}$ [8]. With this choice, the average inter-particle distance $1/\sqrt[3]{n_{j0}}$ becomes smaller than the thermal de Broglie wavelength $\lambda_B$ (i.e., $\lambda_B/\sqrt[3]{n_{j0}} \sim 2.5 > 1$), which means that the quantum effects may no longer be negligible. To investigate the propagation behaviors of the surface plasma waves by the effects of different quantum forces in electron-hole plasmas we analyze Eq. (19) numerically. Figure 2 show the plots of the wave frequency with the wave number K in spin-polarized and unpolarized cases. It is found that the wave frequency increases with increasing values of the hole number density in both the limiting cases of fully (panel a) and zero (panel b) spin polarizations. Physically, lower concentration of the hole species supports low phase speed surface waves in electron-hole semiconductor plasmas.

Figure 3 shows the comparisons of different quantum effects (namely, those associated with the quantum Bohm potential, CE interaction and the degenerate pressure) on the wave frequency in low (panels a and b) and high (panels c and d) density plasmas with fully (panels a,c) and zero (panels b,d) spin polarized plasmas, keeping the effective mass ratio fixed. In both the cases of fully spin polarized plasmas (panels a and c), the effect of the CE interaction is to decrease the wave frequency. It is seen that the CE interaction is more pronounced when the hole concentration is relatively low. On the other hand, the influence of the Bohm potential for zero spin polarized plasmas becomes higher than that for fully spin polarized system. The effect of the spin polarization is to increase the wave frequency in three different cases as shown in Fig. 4, namely, including all of the quantum effects (panel a), in absence of the CE effect (panel b), and only in the presence of the degenerate pressure (panel c). From Figs. 4(a)-4(c), it is also seen that the frequency gets reduced when the effect of the CE force is considered.

## 4 Conclusions

To conclude, we have briefly described the dispersion properties of surface plasma oscillations in an electron–hole semiconductor plasma with the effects of CE interaction. A general dispersion relation is derived and analyzed in some particular cases of interest. Previous results in this context are also recovered. It is found that the dispersion properties of the surface wave modes are quite distinctive in spin-polarized and unpolarized plasmas. Also, higher the concentration of the hole species larger is the phase velocity of surface waves. Furthermore, the effect of the CE force is to reduce the wave frequency in semiconductor plasmas.

The oblique propagation of surface wave relative to the external magnetic field can be a problem of interest, but is beyond of the scope of the present investigation, and is left for a future work. The present results may be useful for understanding the dispersion properties of surface plasma oscillations that may propagate at the interface of plasma-vacuum medium in electron-hole semiconductor plasmas or in solid density plasmas with number densities $n_{j0} \sim 10^6 m^{-3}$ and the magnetic field strength $B_0 \sim 1\text{-}2T$.

*Acknowledgements M. Shahmansouri acknowledges the financial support of Arak University under research Project No. 96/5834. Also, the constructive suggestions of the reviewers are gratefully acknowledged. A. P. Misra is thankful to UGC-SAP (DRS,Phase III) [with Sanction order No. F.510/3/DRS-III/2015(SAPI)] and UGC-MRP [with F. No.*

*43-539/2014 (SR) and*

*FD Diary No. 3668] for support.*

**References**

[1] Rukhazde A A and Shokri B 1998 *Phys. Scr.* **57** 127

[2] Azarenkov N A, Kondratenko A N and Tyshetsky Y 1999 *Sov. Phys. Tech. Phys.* **44** 1286

[3] Schluter M 1997 *J. Phys. D* **30** L11

[4] Ritchie R H 1963 *Prog. Theor. Phys.* **29** 607

[5] Kaw P K and McBride J B 1970 *Phys. Fluids* **13** 1784

[6] Lazar M, Shukla P K and Smolyakov A 2007 *Phys. Plasmas* **14** 124501

[7] Mohamed B F 2010 *Phys. Scr.* **82** 065502

[8] Misra A P 2011 *Phys. Rev. E* **83** 057401

[9] Niknam A R, Boroujeni S T and Khorashadizadeh S M 2013 *Phys. Plasmas* **20** 122106

[10] Misra A P, Ghosh N K and Shukla P K 2010 *J. Plasma Phys*. **76** 87

[11] Khorashadizadeh S M, Boroujeni S T, Rastbood E and Niknam A R 2012 *Phys. Plasmas* **19** 032109

[12] Zhu J, Zhao H and Qiu M 2013 *Phys. Lett. A* **377** 1736

[13] Tyshetskiy Y O, Vladimirov S V and Kompaneets R 2013 *J. Plasma Phys.* **79** 387

[14] Zhu J 2015 *J. Plasma Phys.* **81** 905810110

[15] Misra A P 2007 *Phys. Plasmas* **14** 064501

[16] Moradi A 2015 *Phys. Plasmas* **22** 014501

[17] Shahmansouri M 2015 *Phys. Plasmas* **22** 092106

[18] Moradi A 2017 *Comm. Theor. Phys.* **67** 317

[19] Khalilpour H 2015 *Phys. Plasmas* **22** 122112

[20] Shahmansouri M, Farokhi,B and Aboltaman R 2017 *Phys. Plasmas* **24** 054505

[21] Shahmansouri M, and Mahmodi Moghadam, M 2017 *Phys. Plasmas* **24** 102107

[22] Andreev P A and Kuzmenkov L S 2016 *Appl. Phys. Lett*. **108** 191605

[23] Trukhanova M I and Andreev P A 2015 *Phys. Plasmas* **22** 022128

[24] Andreev P A and Ivanov A Y 2015 *Phys. Plasmas* **22** 072101

[25] Ren H J, Wu Z W, and Chu P K 2007 *Phys. Plasmas* **14** 062102

[26] Wu Z W, Ren H J, Cao J T, and Chu P K 2008 *Phys. Plasmas* **15** 082103

[27] Li C H, Wu Z W, Ren H J, Yang W H, and Chu P K 2012 *Phys. Plasmas* **19** 122114

[28] Moradi A 2016 *Phys. Plasmas* **23** 044701

[29] Moradi A 2016 *Phys. Plasmas* **23** 074701

[30] Moradi A 2016 *Phys. Lett. A* **380** 2580

[31] Manfredi G 2005 *Fields Inst. Commun.* **46**, 263

[32] Andreev P A 2014 *Ann. Phys.* **350** 198

[33] Shahmansouri M and Misra A P 2016 *Phys. Plasmas* **23** 072105

[34] Misra, A P and Samanta, S 2009 *Phys. Plasmas* **16** 074505

**Figure caption**:

Figure 1: Geometrical structure of semi-bounded semiconductor plasma.

Figure 2: The behavior of normalized wave frequency $W$ with respect to the normalized wavenumber $K$, for (a) fully and (b) zero spin polarized, for different values of the hole density concentration, with $m = 0.12$.

Figure 3: The influence of quantum effects on the normalized wave frequency for low hole concentration ($\delta = 0.1$) with (a) fully (b) zero spin polarized and for high hole concentration ($\delta = 0.7$) with (c) fully (d) zero spin polarized, with $m = 0.12$.

Figure 4: The influence of spin polarization on the wave frequency in the presence of (a) CE interaction, Bohm potential (BP) and degenerate pressure (DP), (b) BP and DP effects, and (c) DP effects.

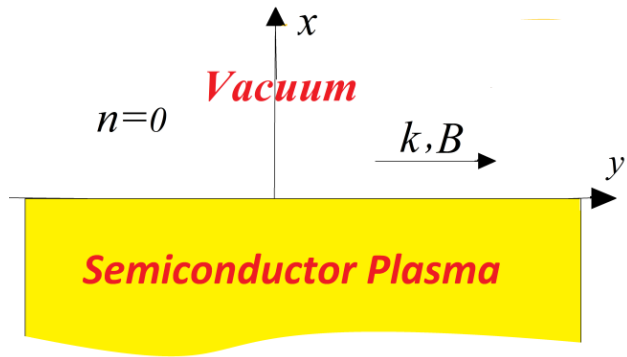

Fig. 1

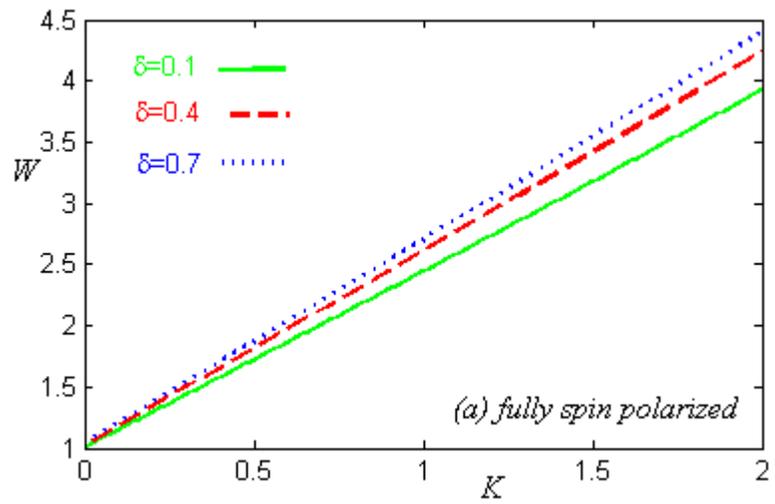

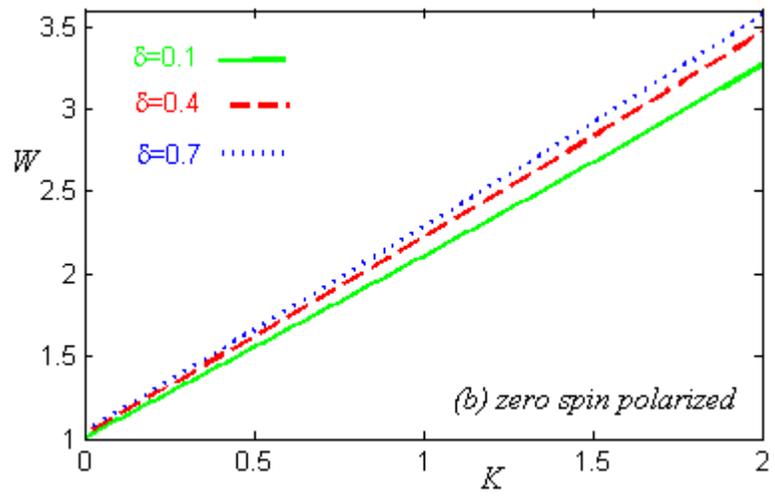

Fig. 2

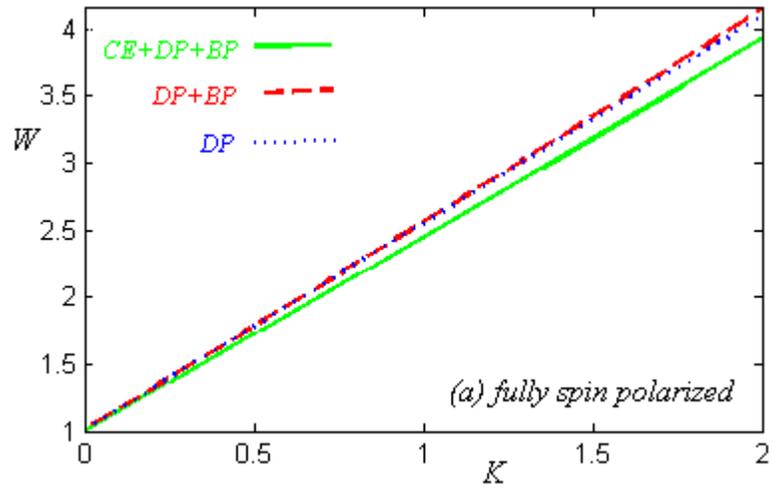

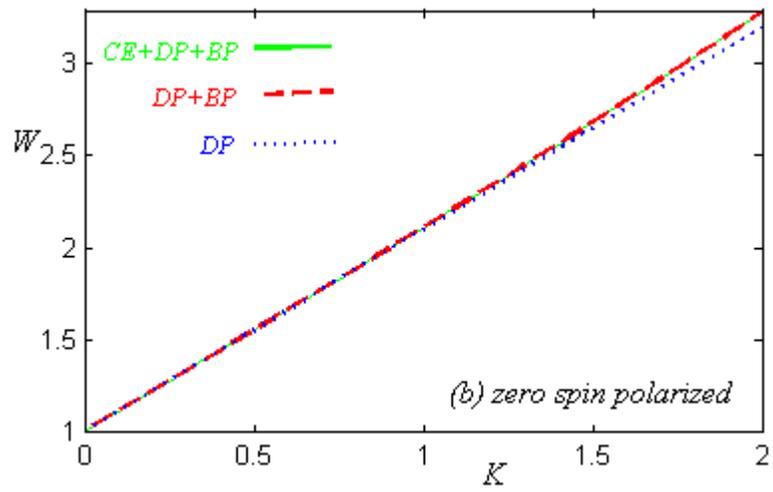

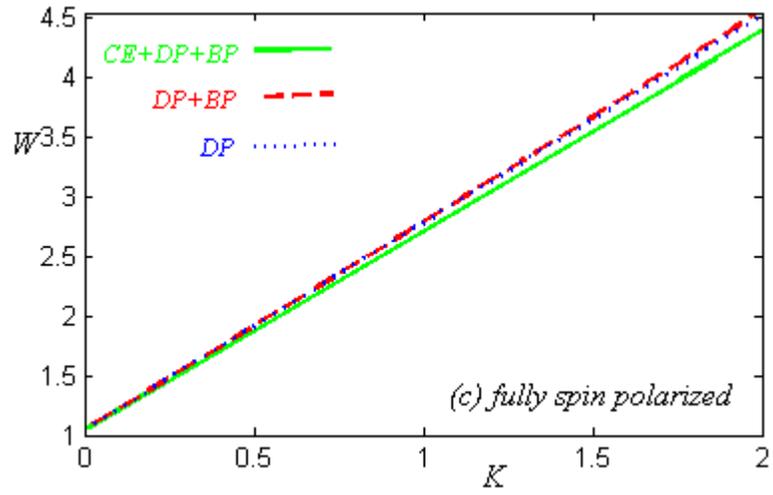

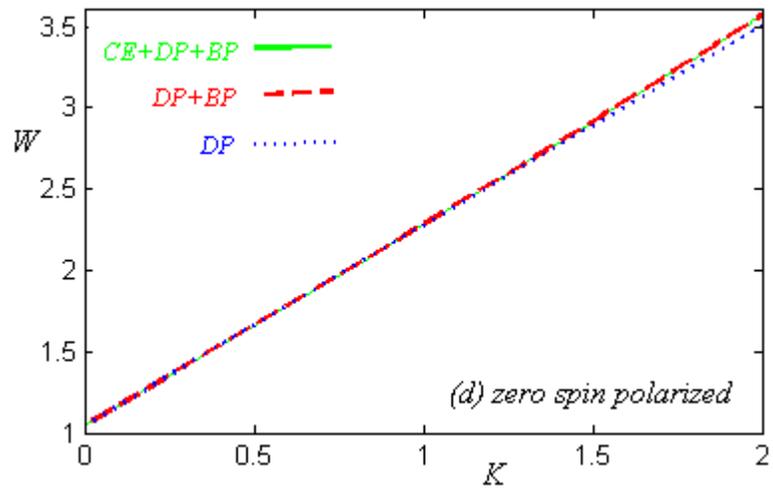

Fig. 3

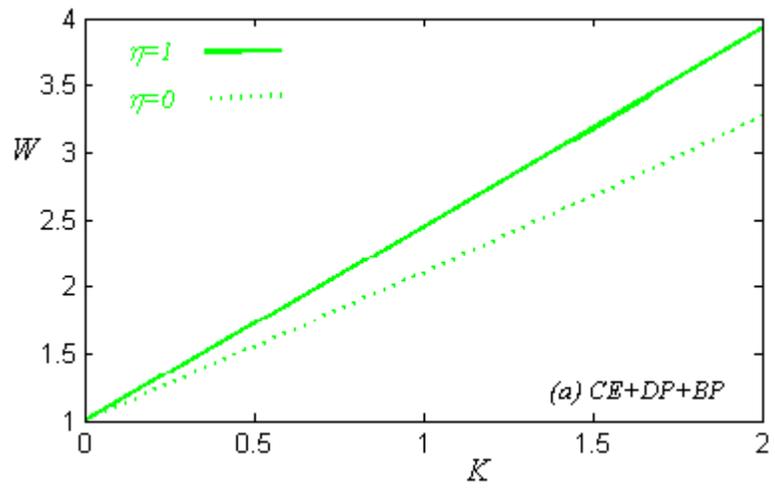

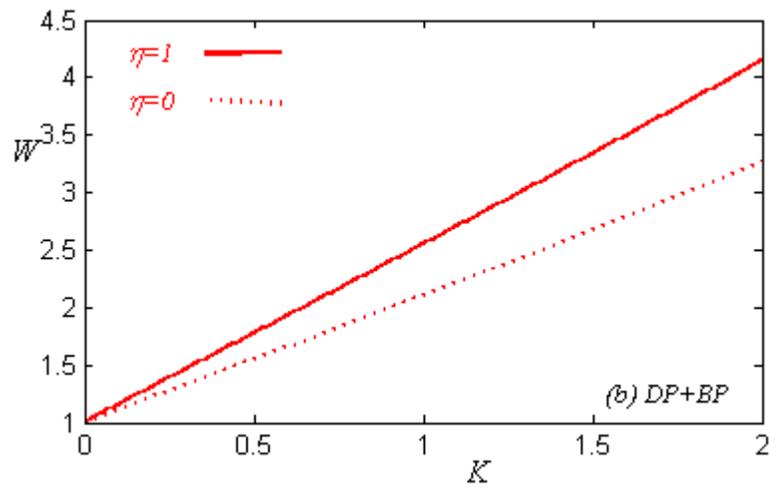

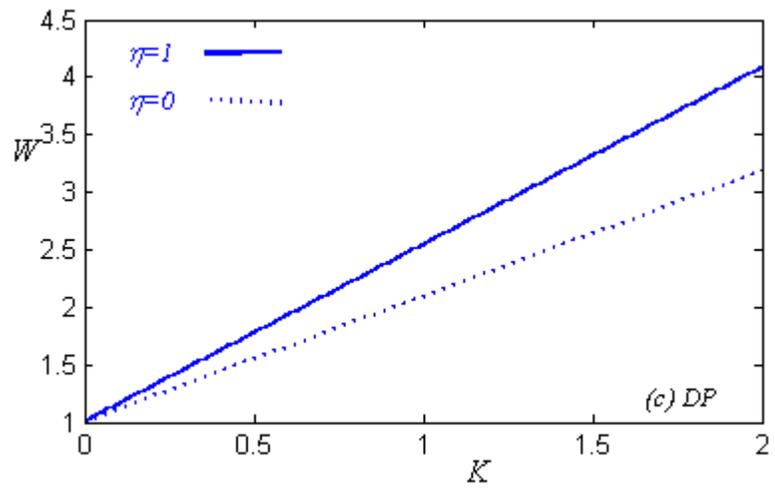

Fig. 4